\documentclass[conference]{IEEEtran}
\IEEEoverridecommandlockouts
\usepackage{cite}
\usepackage{amsmath,amssymb,amsfonts, amsthm}
\usepackage{algorithmic}
\usepackage{graphicx}
\usepackage{textcomp}
\usepackage{xcolor}
\def\BibTeX{{\rm B\kern-.05em{\sc i\kern-.025em b}\kern-.08em
    T\kern-.1667em\lower.7ex\hbox{E}\kern-.125emX}}

\usepackage{microtype}
\usepackage[utf8x]{inputenc}
\usepackage[T1,hyphens]{url}
\usepackage[colorlinks=false]{hyperref}
\usepackage{cleveref}

\usepackage{tabularx}
\usepackage{booktabs}   
\newcommand{\ra}[1]{\renewcommand{\arraystretch}{#1}}

\newcolumntype{R}[1]{>{\raggedleft\let\newline\\\arraybackslash\hspace{0pt}}m{#1}}
\usepackage{threeparttable}
\usepackage{colortbl}

\usepackage{soul}
\usepackage{tikz}
\usetikzlibrary{calc}
\input{anonymous.tex}

\makeatletter
\newcommand*{\rom}[1]{\expandafter\@slowromancap\romannumeral #1@}
\makeatother

\usepackage{enumitem}

\usepackage{subcaption} 
\usepackage{adjustbox}

\usepackage{eurosym}

\begin{document}

\title{The economic value of neighborhoods: Predicting real estate prices from the urban environment}

\author{\IEEEauthorblockN{Marco De Nadai\IEEEauthorrefmark{1}\IEEEauthorrefmark{2} and Bruno Lepri\IEEEauthorrefmark{1}}
\IEEEauthorblockA{\IEEEauthorrefmark{1}FBK,
Trento, Italy}
\IEEEauthorblockA{\IEEEauthorrefmark{2}University of Trento,
Trento, Italy\\
Email: \IEEEauthorrefmark{1}denadai@fbk.eu, \IEEEauthorrefmark{1}lepri@fbk.eu}
}

\maketitle

\begin{abstract}
Housing costs have a significant impact on individuals, families, businesses, and governments. Recently, online companies such as Zillow have developed proprietary systems that provide automated estimates of housing prices without the immediate need of professional appraisers.
Yet, our understanding of what drives the value of houses is very limited. In this paper, we use multiple sources of data to entangle the economic contribution of the neighborhood's characteristics such as \emph{walkability} and security perception. We also develop and release a framework able to now-cast housing prices from Open data, without the need for historical transactions.
Experiments involving $\mathbf{70,000}$ houses in $\mathbf{8}$ Italian cities highlight that the neighborhood's \emph{vitality} and \emph{walkability} seem to drive more than $\mathbf{20\%}$ of the housing value. Moreover, the use of this information improves the nowcast by $\mathbf{60\%}$. Hence, the use of property's surroundings' characteristics can be an invaluable resource to appraise the economic and social value of houses after neighborhood changes and, potentially, anticipate gentrification.
\end{abstract}


\section{Introduction}
The value and the affordability of houses play a vital role for individuals, real estate companies, and local governments. Studies have shown that the real estate accounts for nearly 15\% of the European Gross Domestic Product (GDP)~\cite{ESRB}. Therefore, an accurate prediction of real estate trends and prices may help local governments and companies make informed decisions. On the other hand, housing is one of the largest expenses for most of the people. Thus, the right decision on a house can help them save money, and sometimes make profits for the investment in their homes. 

However, real estate appraisal is a challenging multi-dimensional problem that involves estimating many facets of a property, its neighborhood, and its city. In the real estate market, professional appraisers use Multiple Listing Services (MLS) to estimate prices from similar recently sold properties within the same market area~\cite{pagourtzi2003real}. The recent years have seen the advent of new sources of data and new methods, mainly coming from machine learning~\cite{mcgreal1998} and the computer vision community~\cite{you_image-based_2017,Poursaeed2018}. For example, the availability of historical listings, Open Data, and online markets have allowed companies like Zillow\footnote{\url{https://www.zillow.com/}} and Trulia\footnote{\url{https://www.trulia.com/}} to emerge, and estimate property values from millions of historical listings. Though, these proprietary systems are heavily dependent on the availability of timeless sale transaction data. In addition, the role of neighborhood characteristics on housing value remains not taken into account.

Instead, the neighborhood's dependency on home prices is very well documented. Property infrastructures~\cite{fu2014exploiting}, traffic~\cite{wardrip2011public}, neighborhood popularity and reviews~\cite{fu2014sparse} are found to influence the real estate market.
Pleasant and \emph{walkable} neighborhoods have a direct translation on higher housing prices~\cite{cortright2009walking, washington2018premium}. Security perception and neighborhood environmental physical characteristics impact on economic activity~\cite{Wang:2018:UPC:3184558.3186581}, vitality~\cite{denadai2016death} and liveliness~\cite{de2016safer}, which in turn increase housing prices~\cite{hristova_new_2018}. 
These studies often rely on a limited number of factors, thus neglecting other facets and the complex intricacies inside neighborhoods. This is not to mention the common linearity assumptions between variables under investigation.

In this work, we study how neighborhood's features influence housing values, relating the characteristics of the property together with the environmental, physical, and perceptual characteristics in the surroundings. To do so, we analyze more than $70,000$ online advertisements of the largest Italian real estate website and provide a multi-modal analysis of the price drivers in play. Our model does not require timeless historical listings, and it allows predictions to be responsive to urban changes, thanks to the up-to-date geographical data we use. 

Our experiments show that the neighborhood characteristics have a significant economic impact on housing price. Moreover, the use of this information helps the automatic appraisal of houses, reducing the prediction error of the model by $60\%$. 
The trained model, the data, and the code are also released to the public to foster new research.

The remind of the paper is organized as follows. In the next section we discuss previous work about housing value description and prediction. In \Cref{sec:data} we describe the data we used. \Cref{sec:methods} formalizes how we approach the problem and how we conduct the experiments. \Cref{sec:results} shows our results. Finally, in \Cref{sec:discussion} we discuss the results and present some implications of our work for citizens, local governments, and real estate companies and investors.

\section{Related work}
The task of automatically estimate the market value of houses can be seen as a regression problem, where the price (or the price per square meter) is the dependent variable, while the independent one is the available information that could help to determine correctly the price. Hence, the task can be based on a weighted regression of house features~\cite{pagourtzi2003real}, historical~\cite{tan2017time} and neighborhood prices~\cite{hallac2015network}, but also pictures~\cite{you_image-based_2017,liu_learning_2018}.
For example, You \emph{et al.}~\cite{you_image-based_2017} created a Recurrent Neural Network (RNN) using the images of sold houses in the neighborhood. Liu \emph{et al.}~\cite{liu_learning_2018} combined textual features and external pictures of the sold house to rank and predict the price. Fu \emph{et al.}~\cite{fu2015real} ranked houses through point of interests, their popularity and reviews.
This recent line of research is also of paramount interest for the real estate industry. Let us mention as examples Zillow and Redfin\footnote{\url{http://www.redfin.com}}: these companies collect past sales prices, mortgage records attached to those sales, and prior tax assessments. Then, they relate these variables to the physical features of the property. This allows them to have good estimates for both private users and industry. However, their approach neglect completely the characteristics of the neighborhood. Besides, Zillow and Redfin do not disclose nor release the used data and methods, thus making difficult a clear understanding of their performances.

Researchers also have analyzed some socioeconomic drivers and environmental characteristics that influence the price of houses. 
Cortright \emph{et al.}~\cite{cortright2009walking} found a positive correlation between \emph{walkability} and housing prices in almost all the analyzed US cities. People, indeed, prefer to live in places full of opportunities, and reachable without depending on cars.
This could also be related to the presence of a public transportation system and low traffic~\cite{wardrip2011public}. 
Other researchers found that intangible qualities of neighborhoods like culture~\cite{hristova_new_2018}, perception~\cite{Buonanno2013} and design~\cite{Poursaeed2018} can be related to houses' price. For example, Hristova \emph{et al.}~\cite{hristova_new_2018} found that Flickr tags related to culture are positively associated to urban development and housing prices in a neighborhood, capturing some aspects of the role played by gentrification.
Buonanno \emph{et al.}~\cite{Buonanno2013} combined data from a victimization survey and data from the housing market to estimate the effect of crime perception on the housing prices in Barcelona. Their results show that in districts perceived as less safe than the average, houses are highly discounted.
Poursaeed \emph{et al.}~\cite{Poursaeed2018} found that luxury level and design qualities detected from houses' images are found to impact prices of sold houses, after controlling for the offered price.
Boys \emph{et al.}~\cite{createstreets2017} analyzed six British cities to find that land use, urban form, design, and diversity matter. Moreover, they found that some features such as greenery are not always a positive thing.
Taken all together, however, these works focus on a very limited number of factors per time, neglecting the role played by the others. Moreover, they often assume a linear relationship between the analyzed variables, thus failing in providing a complete and conclusive picture on how the neighborhood influences the housing prices.

For this reason, our work explores many different facets (e.g. \emph{security perception}, the proximity of greenery) of neighborhoods at the same time analyzing geographical data and online advertisements in a non-linear fashion. We also release a scalable Open framework that can be employed by researchers, individuals, and companies to estimate prices without the need of timeless historical data.

\section{Data}
\label{sec:data}
We collected several sources of data for the 8 largest Italian cities: Turin, Milan, Genoa, Bologna, Florence, Rome, Naples, and Palermo. Here we discuss the collected sources of data.

\subsection{Home listings}
We collected online advertisements of the largest Italian real estate website Immobiliare.it\footnote{\url{http://www.immobiliare.it}}. In this website, real estate agencies and private people can upload ads specifying the type and the location of the property they want to sell. Moreover, sellers describe the apartment through a brief description of the dwelling, adding some pictures, and a list of characteristics such as the property type and the square meters, the heating system, the floor number and the maintenance status.

Our data consists of a snapshot of online advertisements temporarily collected on May, 10th 2018 from the website. 

We focus on apartments, attics, detached and semi-detached houses, loft and open spaces that have both geographical coordinates and asked price. We also excluded foreclosure auctions, buildings under construction, and advertisements that are older than one year. The filtered dataset has $73,383$ houses advertised from May, 10th 2017 to May, 10th 2018.

\subsection{Geographical information: OpenStreetMap and Urban Atlas}
We use data from OpenStreetMap\footnote{\url{http://www.openstreetmap.org}}, a collaborative project that allows people to map the environment collectively. With more than 4.3M contributors it is the most valuable source of information publicly available. We downloaded a full snapshot of Italy in March 2018. From this snapshot, we focus on road networks and amenities.

We also make use of the Urban Atlas 2012\footnote{\url{https://land.copernicus.eu/local/urban-atlas}}, a pan-European project that describes the land use and land cover from satellite imagery. With an overall accuracy of $85\%$, it specifies through $20$ classes (e.g., continuous urban fabric, discontinuous low-density urban fabric, green areas, airports) the use of the land.

\subsection{Google Street View images}
We use the urban appearance data temporarily downloaded from Google Street View API\footnote{\url{https://developers.google.com/maps/documentation/streetview}}. First, for each city, we generated a grid of points distant from each other $100$ meters. Then, for each location, we downloaded four 90-degrees images (north, east, south, and west) to have a full panoramic view of the place. The final dataset includes $154,147$ images.

\begin{figure*}[htbp]
\centerline{\includegraphics[width=\textwidth]{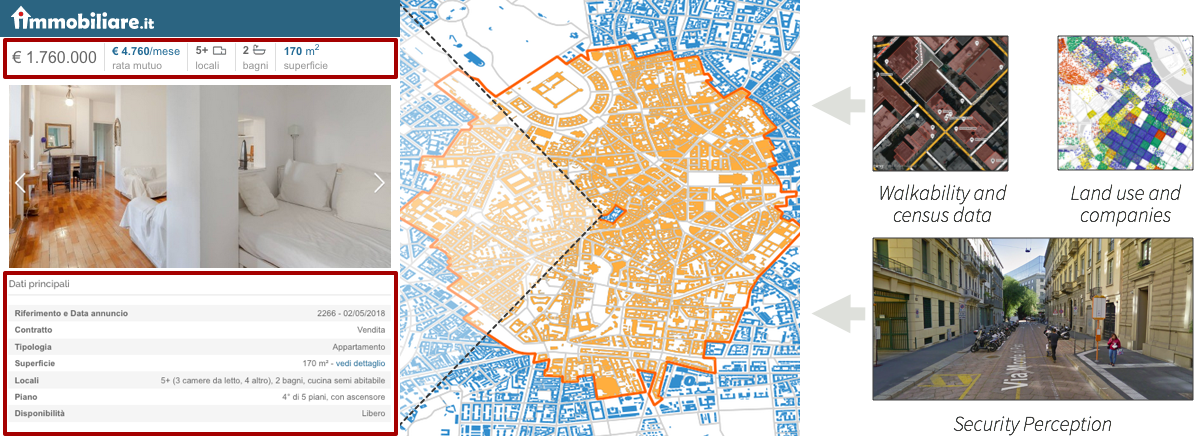}}
\caption{left) An example of housing sales advertisement on Immobiliare.it. We highlighted the information we extract from it; middle) The housing unit is geo-referenced and the neighborhood is built with the census blocks within a 1 km buffer (in orange); right) The model uses the security perception, \emph{walkability}, the data about companies and census of the neighborhood. }
\label{fig:summary}
\end{figure*}

\subsection{Census data and other sources}
The data about population, buildings, and industries are downloaded from the most recent Italian census. The data are publicly provided by the Italian National Institute for Statistics (ISTAT)\footnote{\url{https://www.istat.it/it/archivio/104317}}. We extracted the sizes of businesses from the recent Big Data Challenge 2015\footnote{\url{http://www.telecomitalia.com/tit/it/bigdatachallenge.html}} that released information about companies in 5 Italian cities.

\subsection{Property taxes data}
We also collected property taxes on houses from the disclosed information in advertisements of Immobiliare.it.

\section{Our approach}
\label{sec:methods}

In this paper, we explore the influence played by neighborhood's characteristics on the houses' value. To this end, we leverage multimodal data to obtain (i) security perception scores from Google Street View images, (ii) socioeconomic characteristics from census data, and (iii) the built environment characteristics from geographical Open Data.
First of all, each advertised house is geo-referenced to its containing census block. Then, we compute the features of its neighborhood (e.g., \emph{walkability}, security perception), which are concatenated to the textual characteristics of the property (see \Cref{fig:summary}). We use these data as input to a predictive model that now-casts the value of houses in 8 Italian cities.

\subsection{The neighborhood}
Neighborhoods are the fundamental geographical unit where individuals' activities and social interactions happen the most. Thus, they have always been investigated by social scientists, urban scientists, and criminologists to study human behaviour~\cite{dietz2002estimation}. However, defining the boundaries of a neighborhood is a critical challenge. Usually, they are defined as non-overlapping units through the administrative boundaries defined by the census, but it is unlikely that people move and live obeying to these artificial boundaries. Hence, in our paper we define overlapping neighborhoods resorting to the \emph{egohoods}~\cite{doi:10.1111/1745-9125.12006} of census blocks. 

Starting from a census block, which we will name \emph{ego-place}, we consider as \emph{egohood} all the census areas within a circular buffer of $1$ km. The features of the \emph{egohood} are then computed from this set of blocks (see~\Cref{fig:summary}). 
Thus, we build a spatial binary contiguity matrix $W$ with $W_{i,i} = 0$, where:

\begin{equation*}
  W_{i,j}=\begin{cases}
    1, & \begin{array}{r@{}}
        \text{if $distance(i, j) < 1$ km}\\ 
        \text{and } i \neq j
      \end{array}\\
    0, & \text{otherwise}
  \end{cases}
\end{equation*}
Then, the matrix is row-normalized to have $\sum_j^n W_{i,j} = 1,\quad i=1,2,..,n$.

The \emph{egohood} features $E$ are then computed as the dot product between the spatial matrix and the characteristics of the blocks $F$:
\[
\underbrace{\begin{bmatrix}
x_{0,0}       &   x_{0,1}       & \dots     &   x_{0,c}       \\
x_{1,0}       &   x_{1,1}       & \dots     &   x_{1,c}       \\
\vdots  &  \vdots   &   \vdots  &   \vdots  \\   
x_{n,0}       &   x_{n,1}       & \dots     &   x_{n,c}      \\
            \end{bmatrix}
            }_{\mathrm{\mathbf{place}}-\mathrm{\mathbf{features}}}
\underbrace{\begin{bmatrix}
0      &   1       & \dots     &   1       \\
0       &   0       & \dots     &   0       \\
\vdots  &  \vdots   &   \vdots  &   \vdots  \\   
1       &   0       & \dots     &   1      \\
            \end{bmatrix}
            }_{\mathbf{W}}
    = F \cdot W = E
\]
We will explain in then next sections some precautions to ensure the validity of the hold-out dataset and the cross-validation.

We assume that a house in block $A$ has a value that depends not only by the characteristics of the property and the block but also by the \emph{egohood} that surrounds it. 
As we previously mentioned, the study of the neighborhood effects on human behaviors has a long tradition in social science, economics, urban science and criminology~\cite{dietz2002estimation}.
However, the recent emergence of new sources of data has allowed to empirically connect neighborhood's characteristics with people behaviors at a large scale. In the next sections, we describe the neighborhood characteristics (e.g., \emph{walkability}, \emph{urban fabric}, \emph{cultural capital} and \emph{presence of industries}, \emph{perceived security}, and \emph{living/socioeconomic conditions}) captured by our approach.

\subsubsection{Walkability}
Empirical studies in the US found that neighborhood's \emph{walkability} has a substantial impact on housing prices~\cite{cortright2009walking}. In this report, researchers explored this connection by using the Walk Score index, a proprietary algorithm that scores the walking distance of some typical consumers' destinations. This score ranges from $0$ to $100$, where the first value represents a car-dependent place, while the latter describes a place where all the typical amenities are reachable by foot.
Walk Score accounts for nine different categories of amenities (e.g., restaurants and bars, libraries, grocery stores), and it measures the walking distance through an exponential decay function that reaches $0$ at $1$ mile distance. This score is very convenient for researchers; however, it is proprietary and available only in the US at the present moment. Thus, we here create a similar score based on OpenStreetMap data.

OpenStreetMap contains all the ingredients - the road network, points of interest, and the geographical areas (e.g., parks) - that can be used to create an Open version of Walk Score. Thus, for each \emph{egohood} we compute the \emph{walkability} score
for the following categories: coffee places, entertainment, shopping, restaurants, schools, grocery, library, and parks. 
Similarly to Walk Score, the \emph{walkability} score range is in $(0,1)$. It is computed from a distance decay function that equals $1$ when amenities are less than $500$ meters far, then decays very fast until values are close to a maximum distance $M$, where it starts decaying slowly. Amenities at distances higher than $M$ do not contribute to the score. Thus, the score of block $i$ to amenity $j$ is computed as:
\begin{equation}
  s_{i,j} = e^{- 5 ( d(i, j) /M)^5} 
  \label{eq:walkability}
\end{equation}
where $d(i,j)$ is the number of meters of the shortest path between $i$ and $j$, along the road network. $M$ is the maximum walking distance considered, which is set to $1$. 

Restaurants, bars, shops, and parks are among the most common destinations reachable by walking. In these categories, a variety of options are found to be significant. Thus, the \emph{walkability} score of restaurants is averaged over the ten nearest destinations; this limit is set to five for the shopping destinations, and to two for the park categories. Similar to Walk Score, restaurants and bars are merged into one category due to their overlapping nature.

Public transports represent an efficient and sustainable way to move around the city. For this reason, we also measure the accessibility to the nearest metro and railway stations, the distance to the nearest airport, and the number of bus stops in the neighborhood.

\subsubsection{Urban fabric}
The \emph{vitality} of neighborhoods is believed to depend on how the city is organized and built. Urban activist Jane Jacobs stated that neighborhoods should have four essential conditions to be vital~\cite{jacobs1961death}: \rom{1}) two or more primary uses to have people flocking for different reasons all day long; \rom{2}) small blocks to boost face-to-face interactions; \rom{3}) a mix of old and new buildings to mix big and small enterprises but also people from different income brackets; and \rom{4}) a sufficient concentration of people and enterprises to have a reason to live in the neighborhood.  
Recently, this theory was empirically tested with geographical and mobile phone data~\cite{denadai2016death}. Taking inspiration from De Nadai \emph{et al.}~\cite{denadai2016death}, we here create similar indexes.

From the urban ATLAS dataset we extract the areas dedicated to urban, commercial, and green uses of land. Then, we also use this information to compute the land use mix~\cite{denadai2016death} as:
\begin{equation}
\text{LUM}_i = - \sum_{j=1}^n \frac{P_{i,j} \log(P_{i,j})}{\log(n)}
\label{eq:LUM}
\end{equation}
where $P_{i,j}$ is the probability to have land use $j$ in the \emph{ego-place} $i$. $n=3$ is the number of land uses considered: (i) urban, (ii) commercial, and (iii) green areas.
We also account for the number of residential, commercial and total buildings from census data. Differently from urban ATLAS, these numbers are tight to the census, and thus updated only every ten years. On the contrary, satellite data can be updated very often.

To account for the presence of small blocks, we compute for the average square meters of census blocks. The census institute (ISTAT) delineates these blocks as the smallest enclosed area surrounded by roads or water.

The third Jane Jacobs' condition is about the mixture of companies, old buildings, and people's income. Thus, we extract from census data the "average" size of companies, and the number of buildings for each year-bracket. We also compute the average and standard deviation of the construction years of the buildings.

The fourth and last condition is about residential and commercial density. Thus, we extract the number of companies, shops, and population for each census block.

\subsubsection{Cultural capital and heavy industries}
Cultural capital has been found to influence housing prices and people's behavior in cities~\cite{hristova_new_2018}. Thus, we extract the data of companies with ATECO codes falling in the following macro-categories: 58) Publishing; 59) Film, TV, video, radio, photography; 62-63) IT software and computer services; 71) Architecture; 73) Marketing; 74) Design: product, graphics and fashion; 90) Music, performing, and visual arts; 91) Libraries and museums. Then we add the count of cultural-related companies to the list of features. 

Industries can also influence the health and the likeability of neighborhoods. We account for the presence of heavy industries through the number of heavy industries from census data, and the distance from industrial areas, computed by using OpenStreetMap.

\subsubsection{Security perception} 
Aesthetics is a crucial element to evaluate neighborhood livability~\cite{quercia2014aesthetic}, vitality~\cite{de2016safer} but also social characteristics such as gentrification~\cite{Naik7571}. Moreover, crime perception negatively affects housing prices~\cite{Buonanno2013}. Thus, we estimate the security perception of places through the pre-trained \emph{Safety Perception} Convolutional Neural Network (CNN) model~\cite{de2016safer}. CNNs have recently achieved state-of-the-art performance on visual tasks such
as image and video recognition~\cite{krizhevsky2012,szegedy2015,he2016}. The \emph{Safety Perception} CNN was trained using a crowd-sourced project where people were asked to choose the safer looking place between two randomly chosen Google Street View images. Then, it was fine-tuned with a subset of images from Italy, achieving a $R^2 = 0.62$ over the ground truth labels. Given an image, this model predicts a score in the range $(0, 10)$ where $0$ means that a place is perceived unsafe, while $10$ is considered safe. 
We predict the score of each image from our collected Google Street View dataset, and then we average the scores within the \emph{ego-place}.

\subsubsection{Living conditions}
The cost of housing is very dependent on the living conditions of people. In a city, people's earnings, economic stress, and unemployment can modify the availability of houses, and their price. 
We use the average amount of property taxes as a proxy for the living conditions, to establish a spatial baseline of prices.

\subsection{The property}
The advertisement of the property is provided with a collection of textual features that contain a brief description, and the list of characteristics (e.g., number of rooms, garage, garden). We pre-process these data, and we use a subset of 25 features that are one-hot encoded or transformed to numbers. The considered properties' attributes are: square meters, built year, energy certification, monthly expenses (condominium), floor number, heating type, type of fixtures, garden, furnished, terrace, sun exposition, kitchen type, spa, cellar, garage, fireplace, place type, property class and type, property taxes, condition, and number of rooms, bathrooms, bedrooms. If the advertisement does not mention it, we assume that some features such as cellar, spa, fireplace are absent.

The property is geo-referenced through latitude and longitude information. Thus, we assign the census block that contains the house as its \emph{ego-place}.

\subsection{Model and experimental setting}
In this work, we propose to explore the economic impact of neighborhoods on housing values through a predictive model based on XGBoost~\cite{chen2016xgboost}. It is a widely-used model based on Gradient Boosted Trees, able to scale very well even for high-dimensional data such as the one we analyze. 
Moreover, XGBoost allows to ``open" its black box, thus explaining features' importance and negative/positive contributions of each feature to the predicted values. This answers to the intriguing question ``Why did the model predicted this home value?", a very important answer that helps real estate companies and local governments make important decisions.

The model is defined as nowcasting the value of houses $Y$ from the property's characteristics $P$, and the neighborhood's features $E$ as $P \cdot E \approx Y$.

We train our model using a K-fold cross-validation schema to ensure that the model is robust to unseen neighborhoods and houses. We divide the original dataset into five folds, assigning three to the training set, one to the validation and one to the test set. Then, we iterate the process shifting the folds.
\emph{Now-casting} spatial data poses some challenges on the creation of cross-validation independent folds. Thus, given the train, validation and hold-out sets:
\begin{enumerate}[label={\normalfont(\roman*)}]
\item For each house $x$ in a neighborhood $N_x$ in the training set, it does not exist a house $y$ in the validation (or hold-out) set that is in the same neighborhood of $x$;
\item For each house $z$ in a neighborhood $N_z$ in the validation set, it does not exist a house $u$ in the hold-out set that is in the same neighborhood of $z$.
\end{enumerate}

The model is trained with the following parameters: learning rate of $0.001$, $\lambda=5$, $\alpha=1$, $3$ minimum child weight, max depth of $20$, and $4,000$ estimators, stopping the training when the validation error does not decrease for $50$ rounds. 

We evaluate the results computing the errors on all cross-validation's hold-out sets through the standard Mean Absolute Error (MAE) and the Median Absolute Percentage Error (MdAPE):
\begin{align*}
    \mathrm{MAE} &= \frac{\sum_{i=1}^n\left| y_i-x_i\right|}{n}\\
    p_{i} &= \left|\frac{y_i-x_i}{y_i}\right|\\
    \mathrm{MdAPE} &= \mathrm{median}(\{p_1, p_2, ..., p_n\}) \cdot 100
\end{align*}

\subsection{Open license model}
To the best of our knowledge, there are no released systems that allow to simply deploy, analyze and \emph{nowcast} the real-estate housing market. In this paper, we release a framework able to download, process and predict properties' market prices from textual information and neighborhood characteristics. 
The data used by our Open model comes from heterogeneous sources released with an Open license. 
This potentially allows end-users from universities and research centers, companies, and governments to replicate and update this study at any time. 
This model does not make use of relevant data such as Google Street View images and the property's tax information. For this reason we evaluate it separately in \Cref{tab:jacobsregression}.

\section{Results}
\label{sec:results}

\begin{figure*}
\centering
\adjustbox{valign=b}{
\begin{minipage}[t]{0.48\textwidth}
  \centering
  \includegraphics[width=\linewidth]{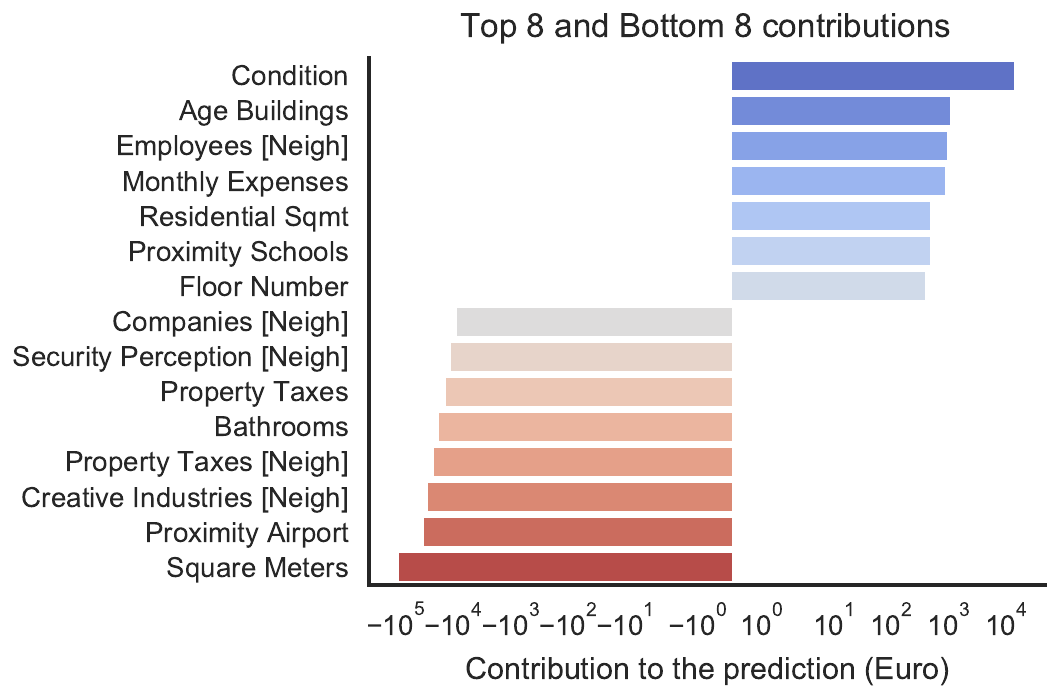}\\
  \vspace{0.2cm}
  (a) Contribution to final prediction
\end{minipage}}\hfill
\adjustbox{valign=b}{\begin{minipage}[t]{0.48\textwidth}
  \centering
  \frame{\includegraphics[width=.49\linewidth]{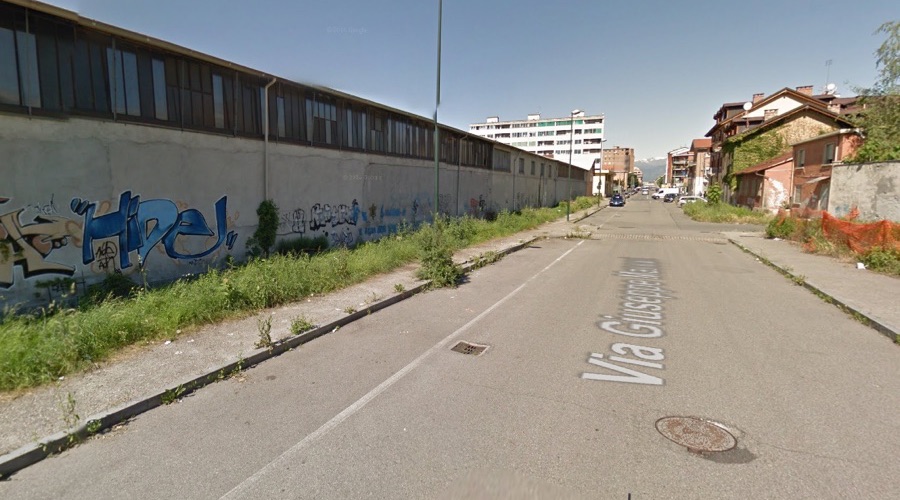}}\hfill
  \frame{\includegraphics[width=.49\linewidth]{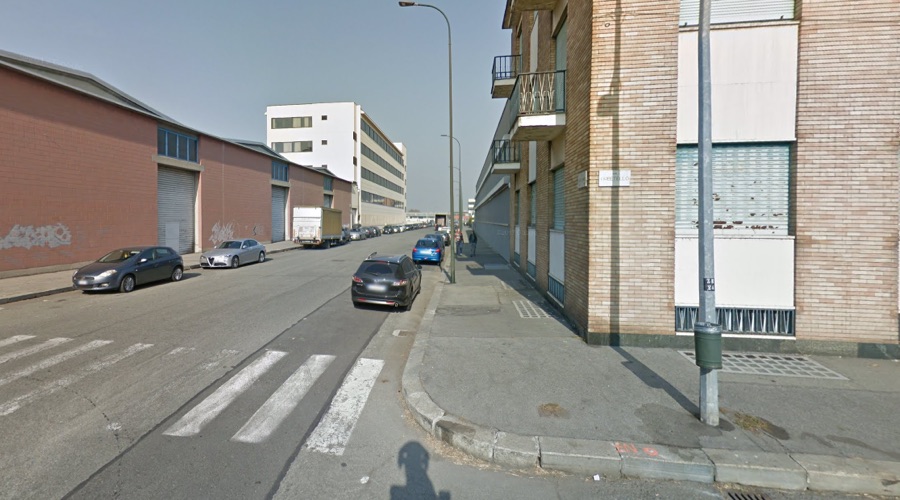}}\\
  \vspace{0.2cm}
  (b) Visual inspection of security perception
  \label{fig:sub2}
\vspace{0.3cm}
{\footnotesize
\begin{tabularx}{\columnwidth}{@{}Xcr@{}}
        \toprule
        \textbf{Feature name} & \textbf{Effect on house's value} & \textbf{Value} \\
        \midrule
        Condition & $\nearrow$& Excellent \\
        Bathrooms       & $\swarrow$& $1$  \\
        Security perception   & $\swarrow$& $3.82$  \\
        Creative industries       & $\swarrow$& $1$  \\
         \bottomrule
    \end{tabularx}}\\
  \vspace{0.2cm}
  (c) Some property's textual features
\end{minipage}}%
\caption{Housing price prediction for a house in Corso Grosseto, Turin. (a) Top positive (negative) contributions of each feature to the final predicted value. Here, the low number of square meters of this house drives the final prediction down. (b) Qualitative valuation of security perception in the \emph{egohood} of the house; (c) Textual features that were very important for this prediction.}
\label{fig:explain}
\end{figure*}

While property's characteristics and market's network effects were widely used to nowcast housing market prices \cite{pagourtzi2003real}, to the best of our knowledge this is the first study that connects the value of houses with its surroundings.

We evaluated our framework for two settings: (i) it uses only textual features of the property, (ii) it uses the characteristics of both the property and the neighborhood to infer the housing price. \Cref{tab:jacobsregression} shows that, among these models, the use of neighborhood features significantly improves the results, dropping the percentage error by $60\%$. 
The $\mathrm{MAE}$ across all cross-validation's folds is around \euro $104,586$, which is very promising accounting that housing prices range from \euro $20,000$ to \euro $20 \text{million}$.
This confirms the important economic role played by the neighborhood, which previously was only hypothesized.

We also tested an \emph{Open} version of the model \rom{2}, that uses only data released with an Open license. This model has an error of $18.02\%$, which is a bit higher than the non-Open model. This is caused by the lack of the \emph{security perception} information and the property taxes of the neighborhood.

\begin{table}[ht!]
    \ra{1.2}
    \caption{The prediction error of real-estate housing prices for three different models. Property uses only textual features of the house (e.g., the number of rooms, floor number). The second model uses both the textual and the \emph{ego-place} features. Its Open version uses only contextual data with an Open license.}
    \centering
    \small
    \begin{tabularx}{\columnwidth}{@{}Xrr@{}}
        \toprule
        \textbf{Model} & \textbf{MAE} & \textbf{MdAPE} \\
        \midrule
        \rom{1}) Property & $148,109$ & $23.78\%$\\
        \rom{2}) Property + Neighborhood   & $\mathbf{104,586}$ & $\mathbf{15.44\%}$ \\
        \rom{3}) Property + Neighborhood (Open)       & $138,929$ & $18.02\%$ \\
         \bottomrule
    \end{tabularx}
    \label{tab:jacobsregression}
\end{table}

\begin{figure}[htbp]
\centerline{\includegraphics[width=\columnwidth]{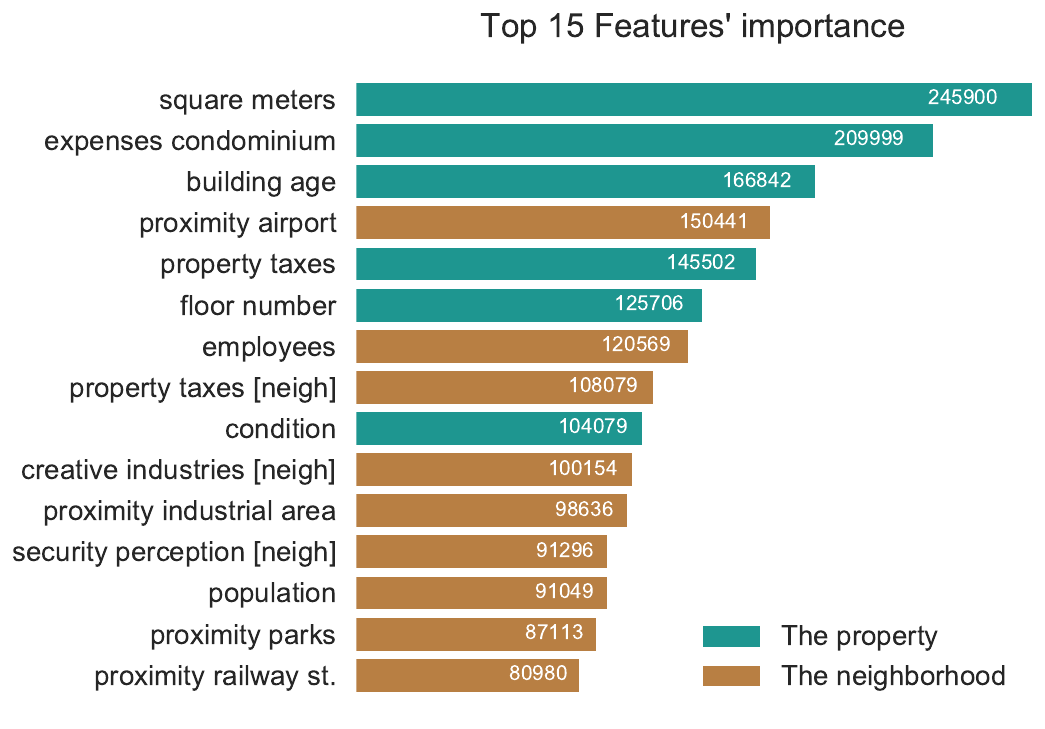}}
\caption{The 15 most important features to predict real-estate housing prices.}
\label{fig:importance}
\end{figure}

The XGBoost model allows us to inspect the most important features among all the predictions. \Cref{fig:importance} shows that square meters, monthly expenses and the age of the building seem to be the primary drivers of price, which is intuitively reasonable. Other notable possible drivers of property's characteristics are its taxes, the floor number and the condition of the apartment. 
However, nine of the 15 top-features are related to the neighborhood. Among them, the number of employees, the population and the proximity of amenities confirms previous studies on the importance of build environment~\cite{denadai2016death}. The presence of creative industries in the \emph{egohood} also seems to have an impact on price, confirming the preliminary results of Hristova \emph{et al.}~\cite{hristova_new_2018} in London and New York. 
It is worth noting that, despite the use of many urban co-variates, visual appearance is one of the most important predictive factors for housing value, as similarly proven for other outcomes such as crime~\cite{Naik7571}, happiness~\cite{quercia2014aesthetic} and presence of people~\cite{de2016safer}.

\subsection{Qualitative/Local Results}
Error metrics and variable importance across all predictions are often not enough to \emph{trust} a model \cite{ribeiro2016should}. To make decisions, especially in policy-making and business settings, end-users have to be confident on how the model reaches a given decision (e.g., the price assigned to the assessed property).   
Often, this may be achieved by using simpler models, with a small number of features, which make people more comfortable with interpreting results. Here, we show how individual predictions are achieved, by breaking down the impact of each feature to the final value of the assessed property.

Tree-based models allow to follow the decision tree to understand \emph{how} each prediction is made. Similarly, we follow each of the XGBoost's boosted tree and then sum the contribution of each variable through the decision path. The final prediction can be the interpreted as the sum of the average bias term and the contribution of each feature:
\begin{equation*}
    y_i = \frac{1}{J}{\sum\limits_{j=1}^J {c_{j}}_{full}}  + \sum\limits_{k=1}^K (\frac{1}{J}\sum\limits_{j=1}^J contrib_j(x, k))
\end{equation*}
where $J$ is the number of trees.

Thanks to this, we now analyze the prediction of a house in corso Grosseto, Turin, from the hold-out set. \Cref{fig:explain} (a) shows the 8 most and least contributing variables in the prediction. The excellent condition of the house increases the price by $12,572$ \euro, while having just one bathroom penalizes the property. 
However, the house is placed in a peripheral and working-class neighborhood of Turin. This emerges from our model. Low-security perception score ($3.82$), the absence of creative industries and low average property taxes significantly decrease the value. For example, it shrinks by $11,713$ \euro, because of the insecure perception of the \emph{egohood}. 
A visual inspection of \Cref{fig:explain} (b) confirms that the lower perception could be related to the presence of graffiti and some disorder typical of the peripheral part of cities.

\begin{figure}[htbp!]
\centerline{\includegraphics[width=\columnwidth]{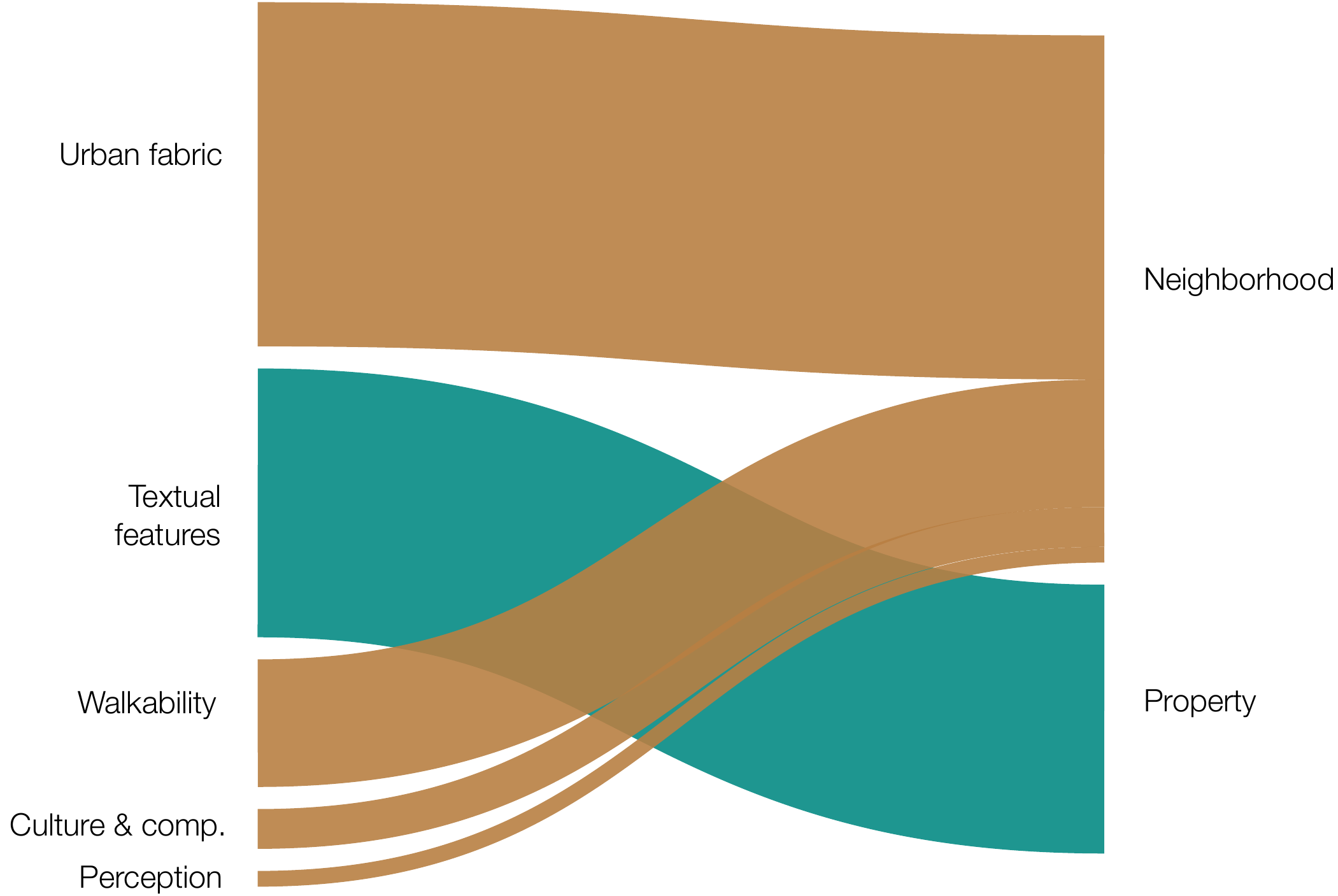}}
\caption{Alluvial plot showing that the features related to neighborhood's characteristics are the most important to predict the housing value.}
\label{fig:alluvial}
\end{figure}

\section{Discussion and implications}
\label{sec:discussion}
Taken together, our results clearly show the economic effect of neighborhood on housing value. As \Cref{fig:alluvial} details, the most important groups of predictors belong to neighborhood metrics, while property's predictors account less. 
Our work is not free of limitations. For example, our collected dataset is prone to noise, as we used advertised houses instead of sold ones. Moreover, we did not observe any temporal trend of the market within cities. This influences how the model learns from one city and test into another.
To the best of our knowledge there are no other papers with code and released datasets. Since we did not have neither one nor the other, we could not apply our work in different countries and compare the diverse baselines we aforementioned. We praise whoever will release geo-located and detailed housing prices, possibly following the guidelines of Loberto \emph{et al.}~\cite{loberto2018potential}.

 However, we believe our paper is the first that quantify the possible economic impact of neighborhood's factors to this extent. Moreover, our machine learning approach relies on geographical and socioeconomic characteristics without the need of historical trades.

From our analyses, we may also draw several implications for citizens, local authorities and urban planners, and for real-estate investors:
\begin{description}
\item \emph{For citizens.} Buying a house is, without any doubt, one of the biggest investments of people life. A right decision can lead to a satisfactory daily living condition without long commuting time to work, with better health outcomes due to higher \emph{walkability} and less traffic, and even higher profits in the future if the house's value increases. An informed judgment on the value of the property is, thus, necessary. People should consider property's characteristics, but also what surrounds the house to buy. Our results found that people should care about the proximity of amenities (e.g., parks, railway stations) but even intangible elements such as the perception of security, which is one of the most important factors. After their investment, people should avoid the vicious loop of degrading, caring about political decisions such as large-scale urban renewal actions, fighting public incivilities such as litter and graffiti, defending neighbors' habits and characteristics. All these factors have an impact on neighborhood's liveliness~\cite{jacobs1961death} and crime levels~\cite{kelling1997fixing}, and thus affect the house prices.
\item \emph{For local authorities and urban planners.} Houses are not islands unto themselves, as they are embedded in complex intricacies of factors in play. Thus, planners have to discourage the creation of oasis in the deserts, where the effect of the neighborhood is demised. For example, \Cref{fig:explain} and \Cref{fig:importance} show that proximity to schools, the presence of amenities and services such as public transport, and absence of creative industries influence housing price. Moreover, changes of the built environment (e.g. the \emph{walkability} and amenities) and spill-over effects of changes in perception can be early signs of gentrification~\cite{Naik7571}. Data and new computational tools represent invaluable tools to analyze cities empirically.
\item \emph{For real-estate investors.} Developers and real-estate investors often think about houses as \emph{units}. This definition should be replaced in favor of \emph{place} and \emph{surroundings}. Investors, especially those working with foreclosures, should examine properties where neighborhoods have high \emph{vitality}, and a virtuous loop of services and improvements over time. 
\Cref{fig:explain} shows that people might prefer to avoid houses near to airports, without creative industries, in neighborhoods where people can not socialize safely.
As Marc Augé argued~\cite{auge2015non}, there are places of every day passing (e.g., malls, airports) where people experience alienation as they cannot live, socialize and have a real identity in the place. He called them \emph{Non-places}. Similarly, neighborhood without the essential social and \emph{vital} qualities can potentially alienate people; hence, they cannot be considered as an adequate investment in this sense.
\end{description}

We released at \url{https://github.com/denadai2/real-estate-neighborhood-prediction} and \href{https://doi.org/10.6084/m9.figshare.6934970}{10.6084/m9.figshare.6934970} the code, trained models, and Open dataset to repeat this study and perform new research.

\section{Conclusion}
In this work, we have studied the economic impact of neighborhood characteristics on housing values. This analysis was formulated as a multi-modal now-casting problem where we predict the price of houses in unseen neighborhoods' conditions.
We collected a large dataset of real estate properties and computed neighborhood features based on overlapping geographical boundaries called \emph{egohoods}.

We trained a Gradient Boosting algorithm on 8 Italian cities, and we found that neighborhood characteristics seem to drive more than $20\%$ of the house's advertised price. Our results also show that the use of this information in the model lowers the prediction error by $60\%$. The qualitative analysis clearly demonstrates the soundness of the proposed solution. We released the code and dataset to foster research in this novel direction.

Indeed, our work suggests several open questions for future work. Can we design experiments and housing policies that can be responsive to neighborhood's changes? Is there a trade-off between urban well-being, the economic success of cities and affordable housing? Can the neighborhood's physical environment predict gentrification? Can open mobile phone dataset, such as Barlacchi \emph{et al.}~\cite{barlacchi2015}, help to predict this dynamic process?  How can we deploy a pipeline able to react to multiple conditions and work adequately in different countries?
We hope that this research, and our released methods, could pave the way for novel studies on the previously neglected link between the cost of houses and the many facets of their surroundings.

\section*{Acknowledgments}
We thank Daniele Foroni and Alberto Danese for the insightful discussions and comments. We also want to thank the Osrm community~\cite{luxen-vetter-2011} for their invaluable routing tool we used in this paper.
The work of Bruno Lepri has been partly supported by the EIT CEDUS Innovation Activity.

\bibliographystyle{IEEEtran}
\bibliography{IEEEabrv,biblio.bib}

\end{document}